\documentclass[preprint,prd,letterpaper,tightenlines,showpacs]{revtex4}
\usepackage{graphicx}
\usepackage{dcolumn}
\usepackage{bm}

\newcommand{\nn}{\nonumber\\}
\newcommand{\la}{\langle}
\newcommand{\ra}{\rangle}

\newcommand{\ben}{\begin{displaymath}}
\newcommand{\een}{\end{displaymath}}
\newcommand{\be}{\begin{equation}}
\newcommand{\ee}{\end{equation}}
\newcommand{\bea}{\begin{eqnarray}}
\newcommand{\eea}{\end{eqnarray}}
\newcommand{\e}[1]{\langle{#1}\rangle}

\newcommand{\eqn}[1]{\label{#1}}
\newcommand{\eq}[1]{Eq.~(\ref{#1})}
\newcommand{\eqs}[1]{Eqs.\ (\ref{#1})}

\newcommand{\bPsi}{\bar{\Psi}}
\newcommand{\bPhi}{\bar{\Phi}}

\newcommand{\he}{\hat{e}}

\newcommand{\uk}{\underline{k}}
\newcommand{\up}{\underline{p}}
\newcommand{\uq}{\underline{q}}
\newcommand{\tG}{\widetilde{G}}
\newcommand{\tGamma}{\Lambda}
\newcommand{\tPsi}{\widetilde{\Psi}}

\newcommand{\K}{K}

\begin{document}

\title{Gauge invariant reduction to the light-front}
\author{A. N. Kvinikhidze}\altaffiliation[On leave from ]{The Mathematical 
Institute of Georgian Academy of Sciences, Tbilisi, Georgia.}
\author{B. Blankleider}
\affiliation{Department of Physics, The Flinders University of South 
Australia, Bedford Park, SA 5042, Australia}
\date{\today}

\begin{abstract}
The problem of constructing gauge invariant currents in terms
of light-cone bound-state wave functions is solved by utilising
the gauging of equations method.  In particular, it is shown how
to construct perturbative expansions
of the electromagnetic current in the light-cone formalism, such that
current conservation is satisfied at each order of the perturbation theory. 

 \end{abstract}
\pacs{}
\maketitle

\section{Introduction}

 Equal light-front``time'' wave functions possess the important
 property of having  boost transformations which are kinematical.
This feature makes the light-cone formalism a 
powerful tool in the investigation of  relativistic processes. 
Lately the light cone approach has often been mentioned  \cite{miller} in relation to
recent measurements of  proton electromagnetic 
form-factors \cite{jones,gayou}.
The light-cone formalism allows one to maintain Poincar\'{e} invariance in a 
simple way, and this can be of great benefit in analysing the
physics behind any particular form-factor behaviour.
In this respect
it would be extremely desirable to develop an approach to the problem
of electromagnetic currents, which combines the three-dimensional nature of 
the boost invariant light-front wave functions, with gauge invariance.
The theory of gauge invariant currents has recently been developed for the usual
four-dimensional Bethe-Salpeter (BS) approach, and for its three-dimensional 
spectator reduction \cite{GrRis,spec}. Here we shall extend this theory to obtain
the gauge invariant three-dimensional reduction to the light-front. 

So far, what has been known \cite {Garsev,Tiburz1} is
that for any given 
two-body BS Green function $G$ and two-body vertex function $\Gamma^\mu$ 
\cite{mandel}, one can
derive a light-front reduced vertex function $\tGamma^\mu$ such that
when sandwiched between light-front wave fuctions [see \eq{lccur}],
 it gives the matrix element of $\Gamma^\mu$ between BS
wave functions [see \eq{BScur}].
The latter is the initial expression for the transition current, and if
it is gauge invariant, then the goal of consructing a gauge invariant
current in terms of the light-cone wave functions is achieved.  
Unfortunately this is not satisfactory from the practical point of view because 
$\tGamma^\mu$ represents an infinite series even in the simplest
case of the one-body Mandelstam current, $\Gamma^\mu=\Gamma^\mu_0$.
In addition, the potential $V$ defining the light-front 
wave function [see \eq{quasipot}], is also only expressible as an infinite series.

The goal of this paper is to derive a conserved current in terms of
light-front bound state wave functions corresponding
to any light-front potential given by equal-time Feynman diagrams.
This enables us to go further, namely, to derive a gauge invariant 
expansion of the current
when only a part of the potential is taken into account nonperturbatively.
Current conservation is satisfied at each order of the perturbation 
theory. 

\section{Light-cone reduction of the two-body equation}
Consider the Green function BS equation for the case of two scalar particles:\footnote{
The case of spinor particles will be considered elsewhere.}
\be
G=G_0+G_0KG.  \eqn{G}
\ee
Define the light-cone two -``time'' Green function 
$\tG(P,\uk,\up)$ as \cite{Garsev}
\be
\tG(P,\uk,\up)= {1\over (2\pi)^2}
\int dk^-dp^-G(P,k,p) \eqn{tilde}
\ee
where the underlined momenta $\up=(p^+, p_\perp)$ denote the
light-cone three-dimensional part of the 4-vector $p=(p^-, p^+, p_\perp)$, 
where $p^\pm=(p^0\pm p^3)/\sqrt{2}$ and $p_\perp=(p^1,p^2)$. There is no
necessity here to specify the precise form for the relative momenta $p$ and $k$;
they could be chosen, for example, as the initial and final momenta of the
second particle.
The equation on the light-cone corresponding to the BS of \eq{G} is
\be
\tG=\tG_0+\tG_0V\tG \eqn{tilG}
\ee
where the light-cone potential is \cite{Garsev}
\begin{eqnarray}\label{quasipot}
V&=&\tG_0^{-1}-\tG^{-1}\nn
&=&
\tG_0^{-1}\left[\e{G_0KG_0}+\e{G_0KG_0KG_0}-
\e{G_0KG_0}\tG_0^{-1}\e{G_0KG_0}+...\right]\tG_0^{-1}.
\end{eqnarray}
Here the angular brakets $\la$ and $\ra$ stand for equating light-cone
``times'' (corresponding to
the integration over relative light-cone energies) in the 
final and initial state, respectively, as in \eq{tilde}. Note that
products of light-cone operators (quantities labelled with a tilde or
enclosed by angular brackets) have implied
three-dimensional integrations over 
$d^3\up=dp^+dp_\perp=P^+dxdp_\perp$ 
in contrast to the four-dimensional integrations implied by products of
BS quantities.

A similar expansion was first derived in \cite{tavkh} (and later rederived in
many papers) for the projection onto the hyperplane where particles have the
same usual time. The infinite series of \eq{quasipot} for the light-cone
potential $V$, suggests that the light-cone wave function be expanded in orders
of the strength of the BS potential $K$. Our task is to construct gauge
invariant currents in terms of these light-cone wave functions.

\section{Gauge invariant currents}

In order to calculate electromagnetic or weak properties of bound states, we
need to construct the corresponding currents. We first start with the BS
approach where the electromagnetic current can be obtained diagramatically by
attaching a photon everywhere in \eq{G} \cite{nnn}.
The resulting expression consists of the matrix element of the vertex function
$\Gamma^\mu$ taken between initial $\Psi\equiv\Psi(P,p)$ and final
$\bPsi\equiv\bPsi(K,k)$
BS bound state wave functions:
\be
J^\mu(K,P)=\bar\Psi\Gamma^\mu\Psi,   \eqn{BScur}
\ee
where
\be
\Gamma^\mu = \Gamma_0^\mu + \K^\mu.
\ee
Here $\Gamma_0^\mu$ denotes the sum of single-particle currents,
and $K^\mu$ is the interaction current. The vertex function 
$\Gamma^\mu$
is related to the gauged Green function $G^\mu$ (five-point function) by
\be
G^\mu=G\Gamma^\mu G \eqn{G-Gam}.
\ee
\eq{BScur} is obtained from \eq{G-Gam} by taking residues at the initial and
final bound state poles \cite{mandel}.
The corresponding axial current can be found in the same way by making an 
axial-vector insertion instead of attaching a photon. 
Define the light-cone two-time five-point Green function $\tG^\mu$ and the
corresponding vertex function $\tGamma^\mu$ by \cite{Garsev}
\be
\tG^\mu(K, \uk; P, \up)={1\over (2\pi)^2}
\int dk^-dp^- G^\mu(K, k; P, p)=\tG\tGamma^\mu\tG.
\eqn{curop}
\ee
Then it's easy to see that the current of \eq{BScur} is also given by a
corresponding matrix element of light-cone quantities:
\be
J^\mu(K,P)= \bar{\tPsi}\tGamma^\mu\tPsi
\eqn{lccur}
\ee
where $\tPsi$ is the light-cone bound state wave function given by
\be
\tPsi(P, \up)=\frac{1}{2\pi}\int dp^- \Psi(P, p).
\ee
The price paid for the relative simplicity of \eq{lccur} (involving the
light-cone wave function $\tPsi$ which depends only on physical
three-dimensional momenta), is the complexity of the light-cone vertex function
$\tGamma^\mu$ which involves an infinite series in powers of $K$ (even for the
case of one-body Mandelstam BS currents, $\Gamma^\mu=\Gamma_0^\mu$),
and the potential $V$ which is also given by an infinite series, \eq{quasipot}.

Clearly, if $\tGamma^\mu$ is given (in all its complexity) by \eq{curop}, and
$\tPsi$ is the bound-state solution of the light-cone bound-state equation
defined by the homogeneous part of \eq{tilG},
\be
\tPsi=\tG_0V\tPsi,
\eqn{tilpsi}
\ee
then the current expressed as the matrix element of \eq{lccur} is conserved, as
it is equal to the matrix element of \eq{BScur}.
However, for practical applications, it is useful to develop a gauge invariant
perturbation theory based on the expansion given by \eq{quasipot}. In this paper
our task is to develop such a theory where gauge invariance is achieved at each
order of the perturbation. By contrast, in a recent series of papers
\cite{Tiburz} this problem has been approached with the strategy of improving
gauge invariance by increasing the order of perturbation.

Our approach is founded on the fact that equating light-cone times
in the initial state ($x_1^+-x_2^+=0$) and similarly in the final state,
which implies integration 
over relative ``energies'' in momentum space, as in \eq{curop}, does 
not change either the Ward-Takahashi identity (WTI) or the Ward identity (WI),
i.e.\
\be
q_\mu\tG^\mu=\he\tG-\tG\he   \eqn{WTI}
\ee
where $q=K-P$ is the momentum transferred by the current to the initial bound
state, and where the operator $\hat{e}$ shifts the momenta and picks up the 
charges of the constituents as required. Its four-dimensional 
form can be found in \cite{nnn}, while in the present light-cone version,
$\he$ is defined by
\begin{eqnarray}
\he(K, \uk; P, \up)&=&i(2\pi)^7\delta^4(K-P-q)
\left[e_1\delta^3(\uk_1- \up_1- \uq)
+e_2\delta^3(\uk_2- \up_2- \uq)\right]\nn
&=&i(2\pi)^7\delta^4(K-P-q)
\left[e_1\delta^3(\uk_2- \up_2)
+e_2\delta^3(\uk_1- \up_1)\right].
\end{eqnarray}
Here $e_i$ (without hat) is the i-th particle charge operator.
We then define the gauging of a two-time quantity as, first the gauging of
the corresponding four-dimensional quantity, and then the equating of times in
the initial and in the final states, e.g. for the Green function we have
\be
\la G \ra^\mu = \left(\tG\right)^\mu=\e{G^\mu}.   \eqn{defgaug}
\ee
It can be argued that \eq{defgaug} is not even a matter of definition, if one
recalls that ``gauging'' is equivalent to taking a functional derivative over an
auxilary field associated with a given current \cite{nnn}, and as such, does not
depend on whether it is taken before or after the times of the particles are
equated.

Using this definition, one can gauge \eq{quasipot}, in this way obtaining
$V^\mu$ expressed as a perturbation series with respect to powers of the
strength of the BS interaction $K$.  A similar perturbation series can be
written for $\tGamma^\mu$ simply from its definition in \eq{curop}. It is then
easy to see that
\be
\tGamma^\mu=\tGamma_0^\mu+V^\mu  \eqn{lccurV}
\ee
where $\tGamma_0^\mu$ is defined in the same way as $\tGamma^\mu$, namely
\be
\tGamma_0^\mu=\tG_0^{-1}\tG_0^\mu\tG_0^{-1}.  \eqn{Lambda_0^mu}
\ee
Note that to obtain \eq{lccurV}, one needs to use the fact that
\be
[\tG_0^{-1}]^\mu = -\tGamma_0^\mu
\ee
which follows from formally gauging the identity operator as 
\be
[\tG_0^{-1}\tG_0]^\mu=\tG_0^{-1}\tG_0^\mu+
[\tG_0^{-1}]^\mu\tG_0=0 .   \eqn{gaugG0}
\ee
The result of \eq{lccurV} is central to this paper, and allows us to
develop the sought-after gauge invariant perturbation theory.

It is evident, that no matter how one defines the perturbation expansion of the
light front potential $V$, each $n$'th order term $V_n$ of the expansion of
\eq{quasipot} 
and the corresponding term $V_n^\mu$ in the expansion of \eq{lccurV}, are
related to each other via the WTI's
\be
q_\mu V_n^\mu= \he V_n- V_n\he .  \eqn{WTIV}
\ee
The current calculated up to $n$'th oder is given by
\be
J_n^\mu = \bar{\tPsi}_n\left(\tGamma_0^\mu + \sum_{i=1}^nV_i^\mu\right)\tPsi_n
\eqn{J_n}
\ee
where $\tPsi_n$ is the corresponding light cone bound state wave function
satisfying
\be
\left(\tG_0^{-1} - \sum_{i=1}^n V_i\right)\tPsi_n = 0.
\ee
Then with the help of \eq{WTIV} and the WTI for $\tGamma_0^\mu$ it is easy to
see that $J_n^\mu$ is conserved.

To give a concrete example of a possible choice for $V_n$, let's define it to be
the $n$-particle exchange contribution in a particle exchange model for $K$.  In
particular, if we write $K=K_1+K_2+\dots$ where $K_1$ is the BS one-particle
exchange term, $K_2$ is the BS crossed two-particle exchange, etc., then, from
\eq{quasipot}, the leading order (LO) contribution to the light front potential
would be given by
\be
V_1=\tG_0^{-1}\e{G_0K_1 G_0}\tG_0^{-1}, \eqn{V_1}
\ee
the next-to leading order (NLO) contribution by
\be
V_2=\tG_0^{-1}
\left[\la G_0 K_2G_0\ra +\la G_0 K_1G_0K_1G_0\ra-
\la G_0 K_1G_0\ra \tG_0^{-1}\la G_0K_1G_0\ra\right]\tG_0^{-1} ,\eqn{cross}
\ee
and so on.  To obtain $V_1^\mu$ we simply gauge \eq{V_1}:
\begin{eqnarray}\label{LOintcur}
V_1^\mu &=&\tG_0^{-1} \e{G_0K_1^\mu G_0} \tG_0^{-1}
-\tGamma_0^\mu \e{G_0K_1 G_0} \tG_0^{-1}
-\tG_0^{-1}\e{G_0K_1 G_0}\tGamma_0^\mu\nn
&+&\tG_0^{-1}\e{G_0^\mu K_1 G_0}\tG_0^{-1}
+\tG_0^{-1}\e{G_0K_1 G_0^\mu}\tG_0^{-1}.
\end{eqnarray}
Note that apart from the term involving $K_1^\mu$ which corresponds to
attachments of a photon inside the kernel, 
$V_1^\mu$ contains attachments to the constituents, e.g. $\e{G_0^\mu K_1 G_0}$.
The two terms with a negative sign can be thought of as subtractions 
to the last two terms. These subtract the contributions of the 
intermediate states of the constituents whose times are equal to each
other, as the latter are exposed in the gauged two-time
free Green function $\tGamma_0^\mu$. This can be seen by using \eq{Lambda_0^mu}
and noting that with the help of \eq{bstat},
the subtraction terms can be replaced in \eq{LO} by one-body currents:
\be-\tGamma_0^\mu \e{G_0K_1 G_0} \tG_0^{-1}
-\tG_0^{-1}\e{G_0K_1 G_0}\tGamma_0^\mu =
-\tG_0^{-1}\tG_0^\mu V_1 -
 V_1 \tG_0^\mu\tG_0^{-1}\rightarrow -2\tGamma_0^\mu.
\ee
The current in LO is then
\be
J_1^\mu=\bPhi\left(\tGamma_0^\mu+V_1^\mu\right)\Phi
\eqn{LO}   
\ee
where $\Phi$ is the solution of the light-cone bound state equation 
in LO:
\be
\left(\tG_0^{-1}-V_1\right)\Phi=0.  \eqn{bstat}
\ee
Conservation of the LO current of \eq{LO} follows from the
WTI's for $V_1^\mu$ [\eq{WTIV}] and the one body vertex function
$\tGamma_0^\mu$, and the equation for the bound state, \eq{bstat}.

To conclude this section it is interesting to compare our prescription 
for constructing the LO current by
gauging the LO light-cone potential [\eqs{lccurV}, (\ref{WTIV}), and 
(\ref{LOintcur})], with the related results of Refs.\
\cite{wallace,phillips} for the case of the usual equal-time
 quasipotential approach.  Using the gauging of equations method,
it was shown in Refs.\ \cite{wallace,phillips} that in order 
to obtain a gauge invariant  transition current, both the 
quasipotential and the electromagnetic current operator should be 
truncated at the same order of the coupling constant. 
Thus in Refs.\ \cite{wallace,phillips}, the construction of
 the gauge invariant approximate 
current involves expansions of  both the four-dimensional five-point Green 
function and the quasipotential, whereas we only need
the light-front potential  given as part of a series expansion [\eq{quasipot}].
Gauging just this part (viz.\ the light-front potential), we derive 
the gauge invariant approximate current. This is a nice but formal 
feature of our approach. A more important difference lies in the fact that
the boost transformation of the usual equal-time wave functions is dynamical, i.e.\
it depends on the interaction \cite{wallace,phillips}. Our gauge invariant 
light-front reduction offers wave functions which depend only on
three-dimensional momenta, they have kinematical boost transformations, and 
provide gauge invariant currents, all at the same time. 
It is a difficult task to construct the approximate gauge invariant currents 
in terms of the covariant wave functions projected onto the hyperplane 
$P\cdot(x_1-x_2)=0$ \cite{Garsev,wallace,phillips}. 
One of the ways for this to be achieved would be in a modification of our 
gauging prescription for such projected Green functions.

\subsection{Currents in the NLO}
Above, we have formally solved the problem of constructing conserved light cone
equal-time currents up to any order in the interaction. In this subsection we
would like to apply our formalism to the case where only the LO term of $V$ is
taken into account exactly, with all higher order contributions being
included as a perturbation.

For this purpose, denote the LO contribution to $V$ by $V_1$ (it can be the
single-particle exchange potential discussed above, or it can be defined some
other way), and  the contributions making up the NLO term by $\Delta$:
\be
V =  V_1+\Delta +\ldots.
\ee
Denoting the correction to the wave function $\Phi$ due to $\Delta$ by 
$\delta\Phi$, the following light-cone equation should be satisfied
\be
(\Phi+\delta\Phi)=
\tG_0(V_1+\Delta)(\Phi+\delta\Phi).
\ee
Treating $\Delta$ as a perturbation, and keeping terms that are at most
linear in $\Delta$, the wave function correction $\delta\Phi$ can be 
expressed as \cite{kvin,birse}
\be
\delta\Phi=\left[\tG^b_1\Delta+
\frac{iP^\mu}{4M^2}\left(\bPhi
{\partial\Delta\over\partial P^\mu }\Phi\right)\right]\Phi,
\eqn{wfcor}
\ee
where 
\be
\tG^b_1=\tG_1-{i\Phi\bPhi\over P^2-M^2} \eqn{Gb}
\ee
is the LO Green function $\tG_1$ with the (unperturbed) bound-state pole
subtracted off. 
The second term in \eq{wfcor} is just a wave-function
renormalisation, due to the dependence of $\Delta$ on the total 
momentum $P^\mu$. 
$\tG_1$ satisfies the 
inhomogeneous light-cone equation
\be
\tG_1=\tG_0+\tG_0V_1\tG_1.
\ee
The full linear in $\Delta$ correction to the current matrix element is
\cite{birse} 
\be
\delta J^\mu=
\bPhi\delta \tGamma_1^\mu\Phi+
\bPhi\Delta^\mu\Phi+
\bPhi \tGamma_1^\mu \delta\Phi+
\delta\bPhi \tGamma_1^\mu\Phi.
\eqn{NLOcur}
\ee
where
\be
\tGamma_1^\mu = \tGamma_0^\mu + V_1^\mu.
\ee
The first term stems from the bound state mass correction to the LO
vertex function, that is,
\be
\delta \tGamma_1^\mu = \delta M^2 \frac{\partial
\tGamma_1^\mu}{\partial M^2}
\ee
where \cite{kvin}
\be
\delta M^2=i\bPhi\Delta\Phi.  \eqn{Lorinv}
\ee
The correction to the current, given by \eq{NLOcur}, is conserved by
construction, since the exact current corresponding to the potential
$V_1+\Delta$ is conserved, and so therefore should be the part that is linear in
$\Delta$. Nevertheless, it will be instructive to show this current conservation
explicitly. In this way we will see that the first term in \eq{NLOcur} is
essential for current conservation.

\subsection{Current conservation in the NLO}

Using  the WTI  for $V_1^\mu$ given in \eq{WTIV}, and the corresponding
WTI's for the one-body current $\tGamma_0^\mu$ and for $\Delta^\mu$,
 one obtains
\begin{eqnarray}
\lefteqn{q_\mu\delta J^\mu=-\delta M^2\bPhi
{\partial\left(\he\tG_1^{-1}-\tG_1^{-1}\he\right)\over
\partial M^2}\Phi+
\bPhi\left(\he\Delta-\Delta\he\right)\Phi}\hspace{1cm}\nn
&&
-\bPhi\left(\he\tG_1^{-1}-\tG_1^{-1}\he\right)
\delta\Phi-
\delta\bPhi\left(\he\tG_1^{-1}-\tG_1^{-1}\he\right)
\Phi. 
\end{eqnarray}
In the above expression $q=P'-P$ where $P$ and $P'$ are the total initial 
and final momenta, respectively; in this respect, it should be noted
that in each of the summed terms above, all quantities
standing to the right of operator $\he$ have total momentum $P$, while  
those standing to the left of $\he$ have total momentum $P'$.
Exploiting the bound state equations $\tG_1^{-1}\Phi=\bPhi\tG_1^{-1}=0$
and making use of
\eq{wfcor}, the previous equation can be written as
\begin{eqnarray}
&&q_\mu\delta J^\mu=-\delta M^2\bPhi
{\partial\left(\he\tG_1^{-1}-\tG_1^{-1}\he\right)\over
\partial M^2}\Phi+
\bPhi\left(\he\Delta-\Delta\he\right)\Phi\nn
&&
-\bPhi\he\tG_1^{-1}\left[\tG_1^b\Delta+
\frac{iP_\mu}{4M^2}\left(\bPhi
{\partial\Delta\over\partial P_\mu }\Phi\right)\right]\Phi
+\bPhi\left[\Delta \tG_1^b+
\frac{iP'_\mu}{4M^2}\left(\bPhi
{\partial\Delta\over\partial P'_\mu }
\Phi\right)\right]\tG_1^{-1}\he
\Phi. 
\end{eqnarray}
A further application of the bound state equations gives
\bea
\lefteqn{q_\mu\delta J^\mu=-\delta M^2\bPhi
{\partial\left(\he\tG_1^{-1}-\tG_1^{-1}\he\right)\over
\partial M^2}\Phi+\bPhi\left(\he\Delta-\Delta\he\right)
\Phi}\hspace{1cm}\nn
&&-\bPhi\he\tG_1^{-1}\tG_1^b\Delta\Phi+
\bPhi\Delta\tG_1^b\tG_1^{-1}\he
\Phi. 
\eea
One can see that the terms responsible for bound-state wave function
renormalization drop out by themselves, whereas  other terms 
contribute zero to $q_\mu\delta J^\mu$ only as a result of partial 
cancellation between each other. We will see below that the
renormalization terms are important in the charge conservation relation as
they account for the charge flowing in the intermediate states which are
accounted for in $\Delta$. Using \eq{Gb}, it's easy to show that for $P^2=M^2$
(see appendix)
\be
\tG_1^{-1}\tG_1^b=1-i
{\partial\tG_1^{-1}\over\partial M^2}  
\Phi\bPhi,\hspace{1cm}
\tG_1^b\tG_1^{-1}=1-i\Phi\bPhi{
\partial\tG_1^{-1}\over\partial M^{2}},  \eqn{last}
\ee
where the derivative of the inverse Green function 
$N=i\partial\tG_1^{-1}/\partial M^2$ also appears in the
normalization condition for the bound state wave function:
\be
\bPhi N \Phi=1.  \eqn{norm}
\ee
Using these results in \eq{last} one obtains
\be
q_\mu\delta J^\mu=-\delta M^2\bPhi'
{\partial\left(\he\tG_1^{-1}-\tG'^{-1}_1\he\right)\over
\partial M^2}\Phi+(\bPhi'\he N\Phi)(\bPhi\Delta\Phi)-
(\bPhi'\Delta'\Phi')(\bPhi'N'\he
\Phi)
\ee
where we have explicitly indicated with a prime those quantities for which
the total momentum is $P'$, and left unprimed those for which the total momentum
is $P$. The Lorentz invariance of the mass correction, \eq{Lorinv}, then leads
to current conservation
\be
q_\mu\delta J^\mu=0.
\ee

\subsection{Charge conservation in NLO}

For a two-particle bound state, requirement of ``charge conservation'' is given
by the condition
\be
J^\mu(P,P) = 2(e_1+e_2)P^\mu.
\ee
It is straightforward to show that the LO current $J_1^\mu$, given by \eq{LO},
satisfies this condition [as indeed does $J_n^\mu$ of \eq{J_n} for any $n$].
Here we show that exact charge conservation holds also in the case
where $J^\mu$ is calculated to NLO in perturbation theory.
For this purpose, it will be sufficient to show that $n_\mu\delta J^\mu(P,P)=0$
where $n=P/\sqrt{P^2}$ is the unit four-vector along $P$.

We start by using the WI's for $\tGamma_1^\mu$ and $\Delta^\mu$ in 
\eq{NLOcur}. The WI for  $\tGamma_1^\mu$ can be written as
\begin{eqnarray}
&&\tGamma_1^\mu(P, \uk; P, \up)=-[\tG^{-1}_1]^\mu(P, \uk; P, \up)\nn
&&=i\left[
e_2{\partial \tG^{-1}_1(P,\uk,\up)\over\partial k_\mu}+
{\partial \tG^{-1}_1(P,\uk,\up)\over\partial p_\mu}e_2+(e_1+e_2)
{\partial \tG^{-1}_1(P,\uk,\up)\over\partial P_\mu}\right]  \eqn{WI}
\end{eqnarray}
where we have taken the particular choice $\up=\up_2$ and $\uk=\uk_2$ for the
relative variables. The derivation of $n_\mu \delta J^\mu(P,P)=0$ for
the two first terms involving $\partial /\partial k_\mu$ and 
$\partial /\partial p_\mu$ is very similar to the one given above for current 
conservation, therefore, we will consider only the case of the last term
in \eq{WI}.

Any function $F$ of the four-vector $P$ can be considered a function of
$|P|=\sqrt{P^2}$ and any three independent components of $n=P/|P|$. In this
case it's clear that
\be 
n_\mu{\partial F(P)\over\partial P_\mu}=
{\partial F(|P|n)\over\partial |P|},
\ee
so that
\be
n_\mu{\partial \over\partial P_\mu}=
{\partial \over\partial |P|}=2|P|{\partial \over\partial P^2}.
\ee
To determine the contribution of the last term of \eq{WI} to
$n_\mu \delta J^\mu$, we need to consider the contractions
\begin{eqnarray}
&&n_\mu\tGamma_1^\mu(P, \uk; P, \up)=
-n_\mu[\tG^{-1}_1]^\mu(P, \uk; P, \up)\hspace{2mm}
\rightarrow\hspace{2mm} e{\partial \tG^{-1}_1(P, \uk, \up)
\over\partial |P|}\nn
&&
n_\mu\Delta^\mu(P, \uk; P, \up)
\hspace{2mm}\rightarrow \hspace{2mm}- e{\partial \Delta(P, \uk, \up)
\over\partial |P|}
\end{eqnarray}
where $e=e_1+e_2$, and the dependence on $|P|$ is found by writing
$P^\mu=|P|n^\mu$. Then
\begin{eqnarray}
n_\mu\bPhi\tGamma_1^\mu\delta\Phi\hspace{2mm}\rightarrow\hspace{2mm}
2e|P|\bPhi{\partial \tG_1^{-1}\over\partial P^2}
\left[\tG^b\Delta+\frac{i}{2}\left(\bPhi
{\partial\Delta\over\partial P^2}\Phi\right)\right]\Phi,
\end{eqnarray}
which for our purpose needs to be evaluated at $P^2=M^2$.
Using the fact that
\be
\bPhi{\partial \tG^{-1}_1\over \partial M^2} \tG_1^b
=-{i\over 2}\bPhi\left(\bPhi
{\partial^2 \tG_1^{-1}\over (\partial M^2)^2}\Phi\right),
\ee
which follows from the derivation given in the appendix, and the normalisation
condition of \eq{norm}, we obtain that
\begin{eqnarray}
n_\mu\bPhi\tGamma_1^\mu\delta\Phi\hspace{2mm}\rightarrow\hspace{2mm}
-ie M \bPhi\Delta\Phi\left(\bPhi
{\partial^2 \tG_1^{-1}\over (\partial M^2)^2}\Phi\right)
+e M\left(\bPhi
{\partial\Delta\over\partial M^2}\Phi\right), \eqn{1}
\end{eqnarray}
\be
n_\mu\delta M^2\bPhi{\partial\tGamma_1^\mu\over\partial M^2}\Phi
\hspace{2mm}\rightarrow\hspace{2mm}2ie M\bPhi\Delta\Phi\left(\bPhi
{\partial^2 \tG_1^{-1}\over (\partial M^2)^2}\Phi\right),   \eqn{2}
\ee
\be
n_\mu\bPhi\Delta^\mu\Phi\hspace{2mm}\rightarrow\hspace{2mm}-2e M\left(\bPhi
{\partial\Delta\over\partial M^2}\Phi\right).   \eqn{3}
\ee
Using \eq{1}, the corresponding expression for 
$n_\mu\delta\bPhi\tGamma_1^\mu\Phi$,
\eq{2} and \eq{3} in \eq{NLOcur}, one obtains that 
$n_\mu\delta J^\mu=0$.

\section{Conclusions}
The equality of the three-dimensional light-front expression for the current,
\eq{lccur}, and the corresponding four-dimensional BS expression, \eq{BScur},
has been known for a long time \cite{Garsev}. However, this result is not very
practical for calculational purposes as both the light-cone vertex function,
defined by \eq{curop}, and the potential generating the light-cone wave
function, \eq{quasipot}, are represented by infinite series even if the
underlying BS kernel $K$ is simple. In addition, it has so far not been noticed
that between these two operators there is a direct connection (even though they
are given by series); namely, $\tGamma^\mu$ can be obtained from $V$ by the
procedure of gauging if the latter is properly defined in terms of two-time
Green functions.  A simple and natural definition of gauging in this paper is
summarised by \eq{defgaug} and \eq{gaugG0}, and makes the last statement
clear.  Our
definition of gauging enables us to construct the current operator corresponding
to any term of the series of \eq{quasipot}. In particular, we have given the
explicit expression for the gauge invariant current [\eq{LO} and \eq{LOintcur}]
corresponding to the first term of \eq{quasipot}. This expression can be used,
for example, in studies of one-particle exchange models.  Finally, we have shown
how to account perturbatively for the remainder of the terms in \eq{quasipot},
by explicit consruction of the current in the NLO [\eq{NLOcur}]. Close
examination shows that all terms in \eq{NLOcur} are important for current
conservation, which is a result of cancellation between their longitudinal
parts.

The perturbation theory presented in this paper, in particular the expression of
\eq{NLOcur}, could be applied, for example, to the calculation of meson cloud
effects on the electromagnetic form-factors, which are known to be important
\cite{miller}. Although a similar program for the NJL model has been
demonstrated in Ref.\ \cite{birse}, this should also be done in the light-front
approach. In this case, first the NLO potential $\Delta$ should be constructed
to incorporate one meson exchange in all possible ways within the LO model.
Next, such a $\Delta$ should be gauged in order to derive the meson exchange
electromagnetic current operator, $\Delta^\mu$, etc.

The gauge invariant perturbation theory proposed in this note is not
specific to the light-cone approach and can, for example, be 
applied to the spectator approach \cite{GrRis,spec}. The spectator potential
corresponding to the example of \eq{cross} reads
\begin{eqnarray}
V_1&=&K_1 \nn
V_2&=&K_2+K_1G_0K_1-K_1\delta dK_1 \eqn{Gross}
\end{eqnarray}
where $\delta d$ is the product of the single particle propagator $d$ and the
spectator on-mass-shell $\delta$ function. Currents would again be given in LO
by \eq{LO} and in NLO by \eq{NLOcur}, however, rather than gauging the
equal-time light-cone propagators, one would gauge the on-mass-shell propagators
instead \cite{GrRis,spec}.


\appendix
\section{}
Here we derive the following three useful expressions for the case of
$P^2=M^2$,
\be
\tG^{-1}_1\tG^b_1=1-i
{\partial \tG^{-1}_1\over\partial M^2} \Phi\bPhi,
\hspace{1cm}
\tG^b_1 \tG^{-1}_1=1-i\Phi\bPhi{
\partial \tG^{-1}_1\over\partial M^2},   \eqn{A1}
\ee
\be
\tG_1^b{\partial \tG_1^{-1}\over \partial M^2}\Phi
=-{i\over 2}\left(\bPhi{\partial^2 \tG_1^{-1}\over (\partial M^2)^2}
\Phi\right)\Phi. \eqn{A2}
\ee

To carry out the necessary algebra it is useful to introduce the following
notation:
\be
G_P \equiv \tG_1(P,\uk,\up) \hspace{1cm} \left. G_M \equiv \tG_1(P,\uk,\up)\right|_{P^2 = M^2.}
\ee
Note that our bound state wave function $\Phi$ is covariant
and does not depend on $P^2$, as discussed in Ref.\ \cite{kvin}.
Using this notation and the definition of $G_1^b$ given in \eq{Gb}, it follows
that
\be
G^{-1}_PG^b_P=G_P^{-1}\left(G_P-{i\Phi\bPhi\over P^2-M^2}\right)=1-
{iG^{-1}_P\over P^2-M^2}\Phi\bPhi.
\ee
As $G_M^{-1}\Phi=0$,
we obtain the first of the equations in \eq{A1} in the limit $P^2=M^2$;
the second equation follows similarly.
The last of the equations, \eq{A2}, results from the following algebra:
\begin{eqnarray}
\lefteqn{G^b_{M}{\partial G^{-1}_{M}\over \partial M^2}\Phi
={\partial G^b_{M}G^{-1}_{M}\over \partial M^2}\Phi-
{\partial G^b_{M}\over \partial M^2}G^{-1}_{M}\Phi
={\partial G^b_{M}G^{-1}_{M}\over \partial M^2}\Phi
=\left.{\partial G^b_{P}G^{-1}_{P}\over \partial P^2}\right|_{P^2=M^2}\Phi}
\nn[2mm]
&&=\left[{\partial \over \partial P^2}
\left(G_P-{i\Phi\bPhi\over P^2-M^2}\right)
G^{-1}_{P}\right]_{P^2=M^2}\Phi\nn[4mm]
&&=\left[{\partial \over \partial P^2}
\left(1-{i\Phi\bPhi G^{-1}_{P}\over P^2-M^2}\right)
\right]_{P^2=M^2}\Phi=i\Phi\bPhi\left[{G^{-1}_{P}\over (P^2-M^2)^2}-
{1\over (P^2-M^2)}{\partial G^{-1}_{P}\over \partial P^2}
\right]_{P^2=M^2}\Phi\nn[4mm]
&&
=i\Phi\bPhi\left\{{1\over (P^2-M^2)^2}\left[G^{-1}_{M}
+(P^2-M^2){\partial G^{-1}_M\over\partial M^2}
+{(P^2-M^2)^2\over 2}
{\partial^2 G^{-1}_{M}\over (\partial M^2)^2}\right]\right.\nn[4mm]
&&\left.-{1\over (P^2-M^2)}\left[{\partial G^{-1}_M\over\partial M^2}
+(P^2-M^2){\partial^2 G^{-1}_{M}\over (\partial M^2)^2}\right]
+O(P^2-M^2)\right\}_{P^2=M^2}\Phi\nn[4mm]
&&
=-\frac{i}{2}\left(\bPhi{\partial^2 G^{-1}_{M}\over (\partial M^2)^2}
\Phi\right)\Phi.
\end{eqnarray}

\end{document}